\documentclass[lettersize,journal]{IEEEtran}
\usepackage{amsmath,amsfonts}
\usepackage{mdframed}
\usepackage{array}
\usepackage[caption=false,font=normalsize,labelfont=sf,textfont=sf]{subfig}
\usepackage{caption}
\usepackage{textcomp}
\usepackage{booktabs}
\usepackage{multirow}
\usepackage{stfloats}
\usepackage{url}
\usepackage{verbatim}
\usepackage{makecell}
\usepackage{hyperref}

\usepackage{graphicx}
\usepackage{ulem}
\usepackage{float}
\usepackage{cite}
\usepackage{paralist}

\usepackage[linesnumbered,ruled,vlined]{algorithm2e}
\usepackage{hyperref}
\usepackage[capitalise,nameinlink,noabbrev]{cleveref}

\usepackage{xspace}
\newcommand{\tool}{\textsc{ROMAN}\xspace}
\begin{document}

\title{\tool: Reward-Orchestrated Multi-Head Attention Network for Autonomous Driving System Testing}

\author{Jianlei~Chi,~\IEEEmembership{Member,~IEEE,}
    Yuzhen~Wu,
    Jiaxuan~Hou,
    Xiaodong~Zhang,~\IEEEmembership{Member,~IEEE,}
    Ming~Fan,~\IEEEmembership{Member,~IEEE,}
    Suhui~Sun,
    Weijun~Dai,
    Bo~Li,
    Jianguo~Sun,~\IEEEmembership{Member,~IEEE,}
    and~Jun~Sun,~\IEEEmembership{Member,~IEEE}
\thanks{This paper was produced by the IEEE Publication Technology Group. They are in Piscataway, NJ.}
\thanks{Manuscript received April 19, 2021; revised August 16, 2021.}
\thanks{Jianlei Chi and Yuzhen Wu contributed equally to this work.}
    
    \thanks{Jianlei Chi, Jiaxuan Hou, and Jianguo Sun are with the Hangzhou Institute of Technology, Xidian University, Hangzhou 311200, China (e-mail: chijianlei@gmail.com; 24241214909@stu.xidian.edu.cn; jgsun@xidian.edu.cn).}
    
    \thanks{Yuzhen Wu, Suhui Sun, Weijun Dai, and Bo Li are with the Qingdao Port International Co., Ltd., and also with the Shandong Port Qingdao Port Group Co., Ltd., Qingdao 266011, China (e-mail: YuzhenWu77@hotmail.com; SuhuiSun1@hotmail.com; WeijunDai1@hotmail.com; BoLi181@hotmail.com).}
    
    \thanks{Xiaodong Zhang is with the University of Science and Technology of China, Hefei 230026, China (e-mail: zhangxiaodong@ustc.edu.cn).}

    \thanks{Ming Fan is with Xi'an Jiaotong University, Xi'an 710049, China (e-mail: mingfan@mail.xjtu.edu.cn).}
    
    \thanks{Jun Sun is with the Singapore Management University, Singapore 178903, Singapore (e-mail: junsun@smu.edu.sg).}
    
    \thanks{Corresponding authors: Xiaodong Zhang and Jianguo Sun.}

}

\markboth{Journal of \LaTeX\ Class Files,~Vol.~14, No.~8, August~2021}%
{Shell \MakeLowercase{\textit{et al.}}: A Sample Article Using IEEEtran.cls for IEEE Journals}


\maketitle

\begin{abstract}
Automated Driving System (ADS) acts as the brain of autonomous vehicles, responsible for their safety and efficiency.
Safe deployment requires thorough testing in diverse real-world scenarios and compliance with traffic laws like speed limits, signal obedience, and right-of-way rules.
Violations like running red lights or speeding pose severe safety risks.
However, current testing approaches face significant challenges: \begin{inparaenum}[1)]
\item limited ability to generate complex and high-risk law-breaking scenarios, and
\item failing to account for complex interactions involving multiple vehicles and critical situations.
\end{inparaenum}
To address these challenges, we propose \tool, a novel scenario generation approach for ADS testing that combines a multi-head attention network with a traffic law weighting mechanism.
\tool is designed to generate high-risk violation scenarios to enable more thorough and targeted ADS evaluation.
The multi-head attention mechanism models interactions among vehicles, traffic signals, and other factors.
The traffic law weighting mechanism implements a workflow that leverages an LLM-based risk weighting module to evaluate violations based on the two dimensions of severity and occurrence.
We have evaluated \tool by testing the Baidu Apollo ADS within the CARLA simulation platform and conducting extensive experiments to measure its performance.
Experimental results demonstrate that \tool surpassed state-of-the-art tools ABLE and LawBreaker by achieving 7.91\% higher average violation count than ABLE and 55.96\% higher than LawBreaker, while also maintaining greater scenario diversity.
In addition, only \tool successfully generated violation scenarios for every clause of the input traffic laws, enabling it to identify more high-risk violations than existing approaches.
\end{abstract}

\begin{IEEEkeywords}
Automated Driving System Testing, Multi-head Attention Network, Scenario Generation and Testing.
\end{IEEEkeywords}

\section{Introduction}
\IEEEPARstart{A}{utonomous} Driving Systems (ADSs) place strong emphasis on reducing traffic accidents, yet safety concerns continue to hinder their widespread adoption~\cite{badue2021self}.
Autonomous Vehicles (AVs) must strictly adhere to traffic laws to ensure safety and discipline on the roads~\cite{tang2023survey,lou2022testing}.
In complex urban traffic environments, AVs may violate traffic laws by misinterpreting signals or failing to yield at congested intersections, potentially causing serious safety incidents.
Therefore, testing an ADS against traffic laws is imperative to ensure its safety and reliability.

\begin{figure}[!ht]
\centering
\includegraphics[width=0.7\linewidth,height=5cm]{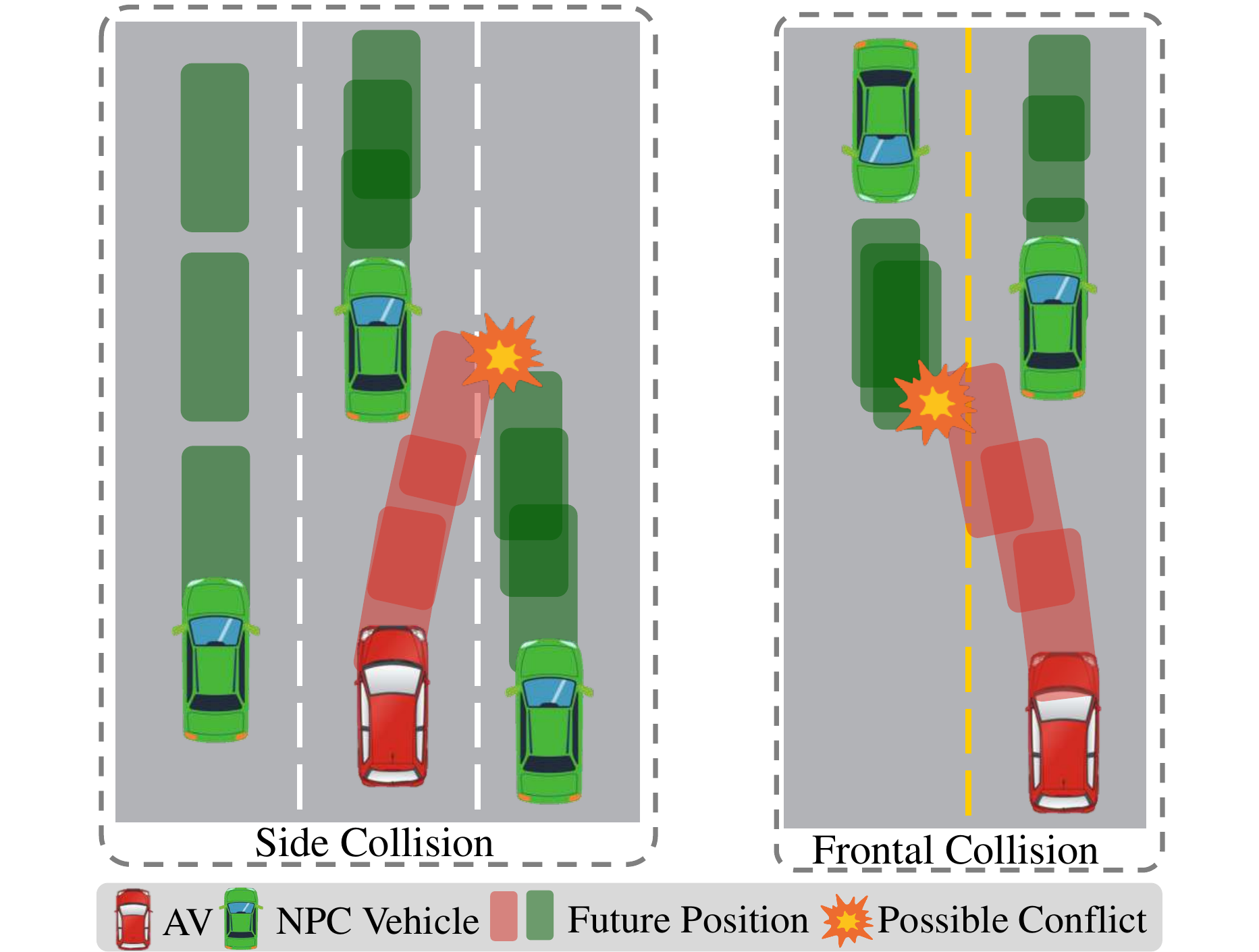}
\caption{Overtaking Violation Scenarios Leading to Collisions}
\label{introduce}
\end{figure}

Identifying high-risk traffic violations is essential when testing an ADS, as it reveals how the system responds to dangerous situations.
The overtaking scenario could serve as an illustrative example of high-risk traffic violations.
A proper overtaking maneuver involves two lane changes, each requiring the AV to signal in advance, maintain a safe distance from vehicles in the target lane, and then execute the lane change appropriately~\cite{streetsurvivalAdvancedDefensive}.
An overtaking violation occurs when the AV fails to correctly perform any of these steps.
As shown in~\cref{introduce}, the AV may misjudge the distance to nearby Non-Player Character (NPC) vehicles, increasing the risk of a side-swipe or even a head-on collision.
Such incidents can result in serious injuries to both drivers and passengers.

To guarantee an ADS's reliability, existing approaches~\cite{zhang2023testing,sun2022lawbreaker
}, e.g., LawBreaker~\cite{sun2022lawbreaker} and ABLE~\cite{zhang2023testing}, have made notable progress in generating law-breaking scenarios when testing an ADS.
LawBreaker is the first framework specifically designed to test traffic law violations in autonomous driving systems.
It employs a genetic-algorithm-guided fuzzing engine to evaluate ADS compliance with diverse traffic regulations.
While LawBreaker effectively uncovers numerous violations in Baidu Apollo~\cite{apollo_v8}, its reliance on a genetic algorithm may lead to premature convergence, reducing the scenario diversity.
ABLE improves on the diversity issue by introducing a GFlowNet-based generative framework~\cite{bengio2021gflow} that samples scenarios in proportion to their likelihood of triggering law violations.
This probabilistic modeling enables broader and more efficient exploration of complex and rare driving situations.
However, ABLE lacks the capability to model rich interactions between multiple agents (e.g., vehicles, pedestrians), limiting its effectiveness in synthesizing multi-vehicle scenarios, an essential aspect of real-world traffic.
Our study shows that ABLE failed to produce any overtaking-related violations even after ten hours of execution.
Other studies also reveal that existing testing methods still struggle to generate diverse, high-risk scenarios with intricate traffic dynamics involving multiple vehicles that capture the complexity of real-world traffic environments.




In response to the aforementioned limitations, we propose an approach \tool, which stands for \textit{Reward-Orchestrated Multi-head Attention
Network}.
\tool is designed to generate diverse, high-risk scenarios and thus thoroughly test an ADS.
It incorporates two novel ideas to guarantee the quality of generated scenarios.
First, \tool leverages a multi-head attention network to model the nuances among multiple diverse environmental factors, such as relative vehicle positions, traffic signals, pedestrian states, and road constraints~\cite{vaswani2017attention}.
This attention-based network effectively captures long-range dependencies, enabling precise modeling of complex interactions in scenarios involving multiple steps~\cite{messaoud2021trajectory}.
Unlike alternative techniques like recurrent neural networks~\cite{rumelhart1986rnn}, multi-head attention excels in processing of multidimensional features during the training phase~\cite{huang2022multi}.
Furthermore, \tool incorporates a novel weighting framework that uses a Large Language Model (LLM) to assess violation risk across the two dimensions of severity and occurrence to enhance the relevance of generated scenarios.
%
This weighting system ensures that violations with significant safety impacts are weighted during scenario generation.
By incorporating these weighted priorities into the scenarios creation process, the algorithm increases the likelihood of high-risk violations occurring during testing, thereby providing a more focused and rigorous measurement for ADS in critical conditions.

We have implemented \tool and integrated it with CARLA, an open-source simulator for autonomous driving~\cite{dosovitskiy2017carla}.
Experimental results show that \tool significantly outperformed ABLE~\cite{zhang2023testing} and LawBreaker~\cite{sun2022lawbreaker}, two state-of-the-art approaches, in law coverage and scenario diversity.
\tool successfully generated violation scenarios for all 81 clauses of the Chinese traffic laws across various road structure types.
Specifically, for intersections with bidirectional roads, ABLE and LawBreaker failed to cover 3 and 8 clauses, respectively.
In this category, \tool achieved a mean violation count of 6.41, exceeding ABLE by 7.91\% and LawBreaker by 55.96\%.
Additionally, \tool reached a high-risk violation rate of 18.25\%, compared to 15.5\% for ABLE and  1.75\% for LawBreaker.
In terms of scenario diversity, which is measured using Dynamic Time Warping (DTW)~\cite{muller2007dynamic} distance, \tool consistently significantly outpaced both ABLE and LawBreaker.
Across all collected DTW distributions, \tool achieved a higher median than the other two tools.
Fundamentally, \tool shifts the paradigm from prior search-based mutation and state-sampling methods to a sequence modeling approach, using a Transformer to learn temporal and structural dependencies within traffic events and holistically compose complex scenarios.

In summary, the main contributions of this work are:
\begin{itemize}
    \item We propose a novel scenario generation approach, \tool, designed to produce diverse, complex, and high-risk scenarios that are more likely to trigger violations of real-world traffic laws.
    \item We employ a multi-head attention network to generate complex ADS scenarios by modeling interactions in traffic scenarios, effectively capturing dynamic interactions.
    \item We introduce a traffic law weighting mechanism that adjusts scenario generation based on the severity and occurrence of violations, with a focus on high-risk violations, and enhance scenario generation by leveraging an LLM to assess and assign risk weights to violations.
    \item We have implemented \tool on the CARLA platform and conducted experiments on the ADS. Experimental results reveal that our approach outperformed existing approaches in key law coverage and scenario diversity. The data and code supporting the findings of this study are openly available. \tool has been publicly released at \footnote{https://anonymous.4open.science/r/ROMAN-6757}.
\end{itemize}

The remainder of this paper is organized as follows.
\cref{sec:Background} defines the formal representation of traffic laws commonly used in ADS testing.
\cref{sec:Approach} details the design of \tool, outlining its core components.
In \cref{sec:Evaluation}, we present our evaluation methodology along with the experimental results.
\cref{sec:Threats} discusses potential threats to the validity of our evaluation.
\cref{sec:Related} reviews two categories of related work, and finally \cref{sec:Conclusion} concludes the paper.
\section{Background}\label{sec:Background}

Traffic laws are foundational to road safety, with many countries imposing strict penalties for serious violations. 
For instance, under Chinese traffic law, a high-risk behavior such as running a red light results in a significant six-point demerit. 
Such official penalties provide a crucial baseline for a violation's severity, but a comprehensive risk assessment must also consider its occurrence in real-world driving.
Therefore, laws associated with severe violations warrant greater attention~\cite{dong2019effectiveness}, and we weight these laws in this paper.

For the development and testing of ADS, compliance with traffic laws is not only a legal requirement but also a critical measure of system safety and reliability. Traffic laws expressed in natural language cannot be directly tested and must be transformed into testable specifications, such as Signal Temporal Logic (STL)~\cite{hekmatnejad2019encoding}. 
We pick up an example from Clause 38, Item 3 of Chinese traffic laws, which has been manually formalized into STL in the prior work of LawBreaker~\cite{sun2022lawbreaker}, to outline the process of formalization.
The clause demands that: \emph{``Right-turning vehicles are permitted to proceed during a red signal, provided that such movement does not impede the passage of other vehicles or pedestrians.''}

The corresponding STL formula is:
\begin{align*}
&G = \{&\\
&\phantom{aa}(traffic\_light\_ahead.color = red \phantom{a} \vee &\\
&\phantom{aaaa}traffic\_light\_ahead.direction.color = red) &\wedge\\
&\phantom{aa}direction = right &\wedge\\
&\phantom{aa}\neg priority\_npc\_ahead &\wedge\\
&\phantom{aa}\neg priority\_peds\_ahead &\wedge\\
&\phantom{aa}(stopline\_ahead(length) \phantom{a} \vee &\\
&\phantom{aaaa}junction\_ahead(length)) &\\
&\}&\\
&G \implies F[0, length](real\_speed > speed) & 
\end{align*}

The condition $G$ (always)  specifies five conditions that must be held at all times, including:
\begin{itemize}
    \item the vehicle is facing either a red traffic light or a red arrow symbol;
    \item the vehicle decides to turn right;
    \item no NPC is crossing in front;
    \item no pedestrian is crossing in front;
    \item the vehicle has stopped before a stop line or junction.
\end{itemize}
Once all five requirements of $G$ are met, the action specified by condition $F$ (eventually) could be executed; the vehicle must accelerate beyond the target speed $speed$ within the distance $length$ to complete the maneuver.
Subsequently, the condition $ F $ is used to evaluate the robustness of the trajectory by ensuring the vehicle achieves the required $speed$ within the specified distance $ length $.

\section{Our Approach}\label{sec:Approach}
\begin{figure}[!htbp]
\centering
\includegraphics[width=0.95\linewidth]{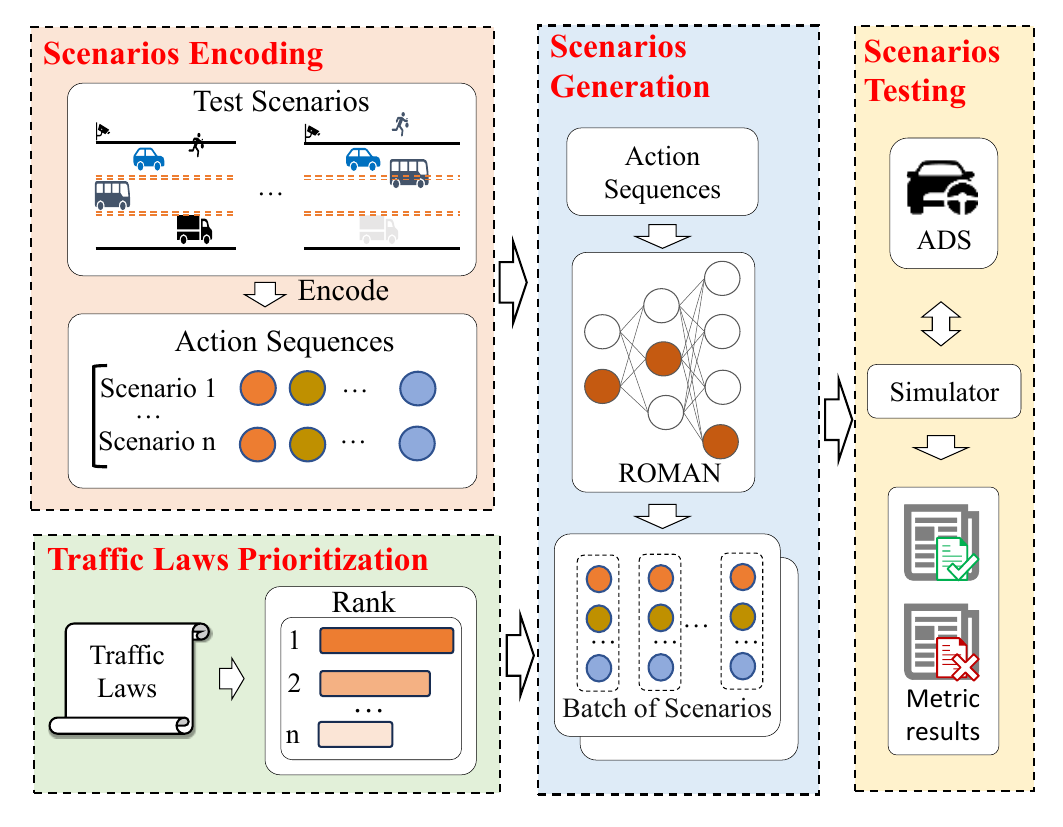}
\caption{The Overview of \tool}
\label{overview}
\end{figure}

In this section, we introduce the architecture of \tool and illustrate all corresponding components.
As a scenario generation approach, \tool takes a set of seed scenarios as input, learns the relationships between scenarios and environmental conditions, and generates optimized scenarios that are more likely to trigger severe traffic law violations.
\cref{overview} presents a high-level overview of \tool, which comprises four main components: \textit{scenario encoding}, \textit{traffic law weighting}, \textit{scenario generation}, and \textit{scenario testing}.

The four components carry out distinct functionality in \tool, such as
\begin{itemize}
    \item \textbf{Scenario encoding}: transforming vehicle trajectories into standardized action sequences,
    \item \textbf{Traffic law weighting}: quantifying violation risk via an LLM-based weighting of severity and occurrence,
    \item \textbf{Scenario generation}: utilizing the encoded action sequences and weighted traffic law constraints to model dynamic changes in complex situations and capture interactions among traffic elements, thereby constructing a group of optimized scenarios,
    \item \textbf{Scenario testing}: evaluating the ADS in a simulator using the generated scenarios.
\end{itemize}

\subsection{Scenario Encoding}



\cref{alg:encoding} presents the detailed algorithm of scenario encoding.
All generated action sequences are stored in the set \( S_a \).
For each scenario in the input set \( S \), it encodes the scenario script into a standardized action sequence, assigns weights based on traffic law demerit points, computes reward values using these weights and trace, and appends the resulting action sequence and reward to \( S_a \), as shown in~\cref{line:foreach}--~\cref{line:append_results}.
Finally, it returns the weighted action sequence set \( S_a \), ensuring greater emphasis on high-priority laws and high-risk scenarios, as shown in~\cref{line:return}.

\begin{algorithm}
\caption{Scenario Encoding and Traffic Law Weighting}\label{alg:encoding}

\KwIn{$\phi(\Theta)$: Violation formula with weights\\
      $S$: Scenario set with trace $\pi$}
\KwOut{Weighted action sequences $S_a$}

\SetKwFunction{FMain}{EncodingandWeighting}
\SetKwProg{Fn}{Function}{:}{}
\SetKwInOut{Input}{Input}\SetKwInOut{Output}{Output}

\setcounter{AlgoLine}{0}
\Fn{\FMain{$\phi(\Theta)$, $S$}}{
    $S_a \leftarrow \emptyset$\;\label{line:init_set}

    \ForEach{scenario\ $s\ in\ S$}{\label{line:foreach}

        Encode $s$ into action sequence $A_s$\;\label{line:encode_scenario}

        Calculate risk weight $w_s$ via LLM-based  weighting of severity and occurrence\;\label{line:assign_weights}

        Compute reward\ $R_s$ as the highest STL robustness from $\pi$ for $\phi(\Theta)$, weighted by $w_s$\;\label{line:generate_rewards}

        Append $A_s$ and $R_s$ to result set $S_a$\; \label{line:append_results}
    }

    \Return $S_a$\; \label{line:return}
}
\end{algorithm}

\begin{table*}[!htbp]
\centering
\caption{Traffic Scenario Encoding Examples}
\label{tab:encoding}
\small
\begin{tabular}{llll}
\toprule
\textbf{Element} & \textbf{Encoding Format} & \textbf{Example Input} & \textbf{Encoding Output} \\
\midrule
Time & \normalfont{time+\textless hour\textgreater} & {\{hour:8,minute:2\}} & \normalfont{time+8+2} \\
\midrule
Weather & \normalfont{weather+\textless type\textgreater+\textless value\textgreater} & {\{rain:0.23\}}  & \normalfont{weather+rain+0.23} \\
\midrule
\multicolumn{1}{l}{\multirow{2}{*}{Ego Vehicle}} & \normalfont{ego+\textless lane\textgreater+\textless offset\textgreater} & {\{lane:``lane1",offset:0.5\}} & \normalfont{ego+lane1+0.5} \\
 & \normalfont{ego+speed+\textless speed\textgreater} &{\{speed:5.3\}} & \normalfont{ego+speed+5.3} \\
\midrule
\multicolumn{1}{l}{\multirow{2}{*}{NPC}} & \normalfont{\textless npcid\textgreater+\textless lane\textgreater+\textless offset\textgreater} & {npc1,\{lane:``lane3",offset:1.5\}} & \normalfont{npc1+lane3+1.5} \\
 & \normalfont{\textless npcid\textgreater+\textless speed\textgreater} & {npc1,\{speed:3.7\}} & \normalfont{npc1+speed+3.7} \\
\bottomrule
\end{tabular}
\end{table*}

In this phase, \tool encodes raw traffic scenarios into standardized action sequences.
The original scenario scripts may apply varying JSON formats, and these structural inconsistencies hinder input processing and complicate subsequent scenario generation.
To resolve such issues, we unify the raw data into standardized action sequences through schema normalization.
As shown in~\cref{eq:encoded_sequence}, a standardized action sequence is a list of \(a'_i\), each denoting an encoded action unit.
\begin{equation}
A_{\text{s}} = (a'_1, a'_2, \dots, a'_L)
\label{eq:encoded_sequence}
\end{equation}

\cref{tab:encoding} lists action unit instances across different categories.
For example, time data like \texttt{\{hour:8, minute:2\}} is encoded as ``\texttt{time+8+2}''.
Weather data, like \texttt{\{rain:0.23\}}, becomes ``\texttt{weather} \texttt{+rain+0.23}''.
Similarly, the ego vehicle's state, e.g., \texttt{\{lane:`lane1', offset:0.5\}} and \texttt{\{speed:5.3\}}, is represented as ``\texttt{ego+lane1+0.5}'' and ``\texttt{ego+speed+5.3}'', respectively.
Notably, while \tool encodes the initial states of the ego vehicle to initialize the scenario, the vehicle remains under the continuous control of the target ADS during testing to evaluate its law obedience.
This approach reduces computational complexity while maintaining the semantic integrity of the original traffic scenario script. The encoding preserves spatial attributes such as location and direction through lane offsets and velocity vectors, while accounting for traffic law requirements under varying visibility, particularly the specific lighting regulations mandated for daytime and nighttime operations.
The resulting encoded action sequence is subsequently used to train the multi-head attention network.

\subsection{Traffic Law Weighting}
\begin{figure}[!htbp]
\centering
\includegraphics[width=1\linewidth]{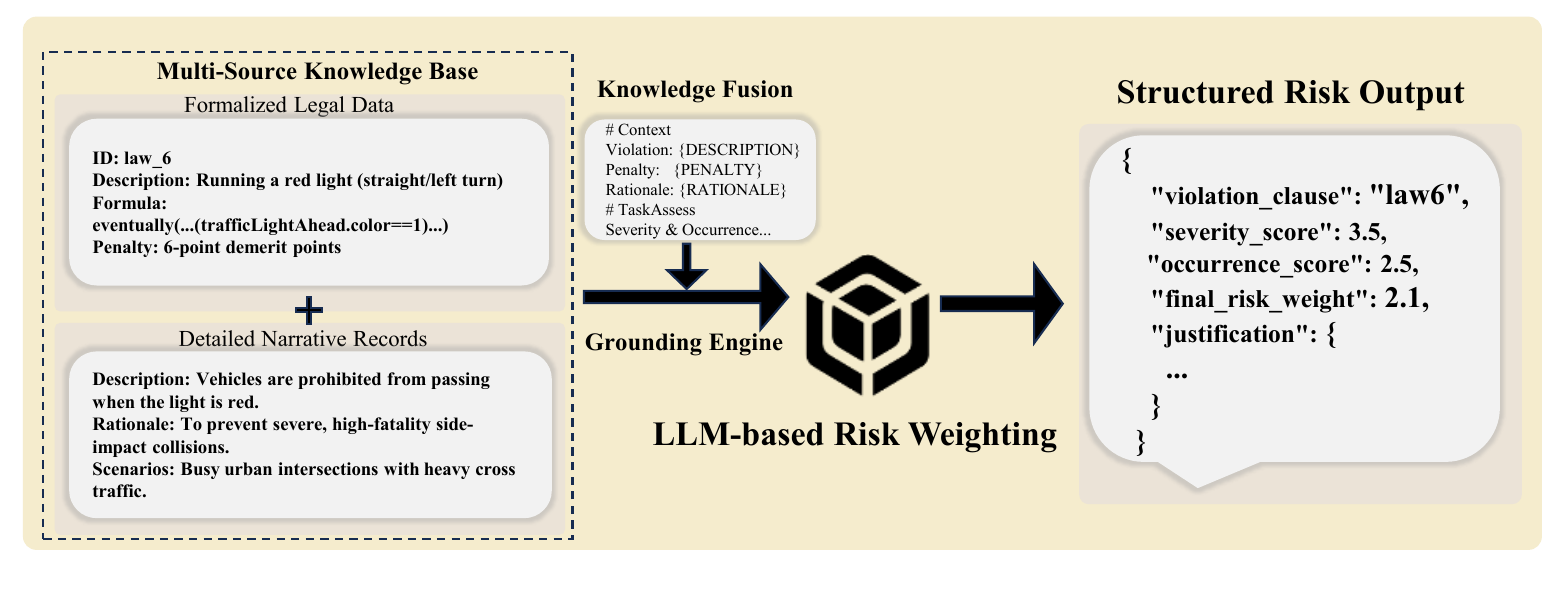}
\caption{Automated Workflow for LLM-based Risk Weighting}
\label{llm_dialogue}
\end{figure}

In this phase, \tool adopts an automated workflow to assess the risk of traffic law violations in a structured and systematic way, as illustrated in~\cref{llm_dialogue}.
The workflow leverages an LLM-based risk weighting module that evaluates violations in terms of both severity and occurrence.
The LLM is guided by a multi-source knowledge base that integrates official documents, including the National Penalty Regulation for Traffic Violations~\cite{Ministry_of_Public_Security_2021} and the National Driver Training and Safety Manual~\cite{MOT_MPS_2022}.
The knowledge base is first constructed by processing raw legal and safety texts into two structured formats.
These consist of formalized legal data containing technical elements like identifiers, penalties, and logical formulas, and detailed narrative records providing narrative context such as causes, scenarios, and descriptions in natural language.
To maintain accuracy, the system performs internal consistency checks that link related records across different data sources, ensuring that all descriptions, penalties, and definitions remain aligned.
For each violation, the grounding engine retrieves both formalized legal data and narrative records from the knowledge base using the violation’s unique identifier.
This information is then systematically assembled into a structured prompt containing legal identifiers, narrative descriptions, and standardized scoring criteria.
This mechanism constrains the LLM, compelling it to ground its assessment on the provided evidence, which includes technical formulas, penalties, and narrative descriptions.
The workflow then produces a structured output containing violation clauses, scores, justifications, and the final risk weight, creating an evidence-based framework that can be applied across regions.
Specifically, the framework remains independent of underlying legal specifications. By expressing traffic regulations from different countries in STL, \tool achieves seamless portability across international jurisdictions while preserving its core architecture.

The LLM evaluates each violation $\phi_{i}$ by assigning a severity score $S(\phi_{i})$ and an occurrence score $O(\phi_{i})$ on a standardized 0.0 to 4.0 scale.
We adopt this range to align with safety engineering conventions, such as ISO 26262-style hazard analysis~\cite{iso26262}, for consistent categorical risk assessment.
This scale was selected to provide clear differentiation between low and high risks while remaining simple and consistent, with descriptive anchors such as 4.0 defined as catastrophic or fatal to guide the LLM’s judgments.
Through this process, the model grounds its reasoning in authoritative sources, improving the relevance and consistency of generated violation scenarios, as shown in~\cref{risk_weight}.
\begin{equation}
\label{risk_weight}
w(\phi_{i}) = (S(\phi_{i}) \times O(\phi_{i})) / 4.0
\end{equation}
To ensure the objectivity and professional correctness of the generated weights, we implement an automated filtering mechanism to perform syntactic and semantic consistency checks on LLM outputs, followed by a comprehensive validation protocol. This process incorporates an expert review conducted by a diverse team including one senior traffic police, two ADS developers, and two test engineers to align weights with real-world legal enforcement.
Furthermore, we cross-validated the weighting results across several representative LLMs, such as GPT, Gemini, Claude, and Qwen, to mitigate model-specific biases and ensure the orchestrated rewards reflect a broad technological consensus.
The resulting weight $w(\phi_{i})$ is then utilized for reward computation in scenario generation.

These weights allow \tool to focus on high-risk violations by adjusting rewards based on the robustness degree of STL formulas $ \phi_i \in \phi(\Theta) $ over the trace $ \pi $.
Here, $\phi(\Theta)$ is the set of relevant violation formula.
For each law $\phi_i$, we compute its robustness $r_{\text{STL}}(\phi_i, \pi, t)$ using STL quantitative semantics~\cite{rodionova2020safe}.
This measures how far the trace $\pi$ deviates from satisfying the STL formula $\phi_i$ at time $t$.
\tool assigns a weight $w(\phi_i)$ to each law $\phi_i$, weighting laws based on potential severity.

Based on trace $\pi$ and weights $w(\phi_i)$, the overall scenario reward $R_{\text{overall}}$ is computed in two steps, as shown in~\cref{eq:reward_computation}.

\begin{equation}
\begin{aligned}
R_{\phi_i} &= \max_{t \in \pi} (r_{\text{STL}}(\phi_i, \pi, t) \times w(\phi_i)) \\
R_{\text{overall}} &= \max(\{R_{\phi_i} \mid \phi_i \in \mathcal{F}_{\text{sub}}, R_{\phi_i} \ge 0\})
\end{aligned}
\label{eq:reward_computation}
\end{equation}
where $R_{\phi_i}$ is the individual score for law $\phi_i$, derived from the maximum weighted STL robustness over trace $\pi$.
$R_{\text{overall}}$ is the final reward.
It is the maximum absolute value among negative robustness scores $R_{\phi_i}$ within a subset of formulas $\mathcal{F}_{\text{sub}} \subseteq \phi(\Theta)$.
This maximum-based aggregation focuses the overall reward on the single most critical rule within the evaluated subset.
The computed $R_{\text{overall}}$ serves as the ground truth reward signal.

This reward function weights scenarios that trigger high-weight, high-risk violations, enhancing the testing of critical safety issues in ADS.
The computed reward guides the multi-head attention network to generate high-reward scenarios.
While every violation warrants attention, our weighting mechanism prioritizes the generation and discovery of rare but critical scenarios, considering the finite nature of testing resources.
Minor violations (e.g., minor speeding) are not entirely ignored but are logged and analyzed as part of the overall testing process.

\subsection{Scenario Generation}
This section outlines the scenario generation process, as described in~\cref{alg:scenario-generation}, which employs a training set of action sequences as inputs.
The generation process in \tool is non-iterative at inference time. Iterative optimization is performed only during training using proxy rewards, which enables efficient generation of high-reward scenarios without requiring feedback loops at runtime.
It uses a multi-head attention network to model action sequences, capturing multi-level interactions among elements while generating context-embedded representations.
This process ultimately generates a batch of high-reward scenarios.
Specifically, \tool is trained with proxy model on the weighted training set \( S_a \) as shown in~\cref{line:train_man2}, each action sequence in \( S_a \) is processed by using \tool to model the scenario context and sample a new action sequence accordingly as shown in~\cref{line:foreach2}--~\cref{line:sample_sequence}, and the sampled sequences are decoded into scenarios and returned as the final batch as shown in~\cref{line:decode2}--~\cref{line:return2}.

\begin{algorithm}
\caption{Scenario Generation}
\label{alg:scenario-generation}
\KwIn{$S_a$: Weighted training set of action sequences}
\KwOut{Scenario batch $B$}

\SetKwFunction{FMain}{ScenarioGeneration}
\SetKwProg{Fn}{Function}{:}{}
\SetKwInOut{Input}{Input}
\SetKwInOut{Output}{Output}

\setcounter{AlgoLine}{0}
\Fn{\FMain{$S_a$}}{
    Train \tool on $S_a$ using proxy model\; \label{line:train_man2}

    \ForEach{action sequence $A \in S_a$}{\label{line:foreach2}
        Compute context $C_A$ using \tool \; \label{line:generate_context}

        Sample the action sequence  $\tilde{A}$ from $ C_A $ \; \label{line:sample_sequence}
    }

    Decode the sequence set $\{\tilde{A}\}$ into $B$\; \label{line:decode2}

    \Return $B$\; \label{line:return2}
}
\end{algorithm}

\subsubsection{Multi-Head Attention Mechanism}

In this phase, \tool employs a multi-head attention mechanism to model complex interactions among traffic scenario features. Key elements, including vehicle positions, traffic signals, weather and road constraints, are tokenized and embedded into a high-dimensional feature space.
To explicitly model the complex interactions between vehicles, \tool represents the entire multi-vehicle scenario as a unified sequence of state tokens and leverages a multi-head attention network.
This allows the model to learn the joint features of high-risk scenarios, which are characterized by a specific spatiotemporal combination of multiple vehicle trajectories rather than the dangerous behavior of a single vehicle.
For example, in an overtaking scenario, a high-risk event is formed by a specific combination of the main vehicle accelerating, the overtaken vehicle not slowing down, and the oncoming vehicle approaching at high speed.
This enables \tool to generate logically coherent and inherently interactive trajectory scripts that are difficult to produce using methods based on genetic algorithms or simple sampling.
By doing so, \tool moves beyond modeling isolated behaviors and instead learns the joint probability distribution of multi-vehicle trajectories that are characteristic of high-risk situations.
The multi-head attention mechanism processes these embeddings through multiple attention heads, each capturing distinct representation subspaces and multi-scale patterns of feature interactions.
This enhances the model's expressive capacity to represent intricate relationships, such as vehicle interactions at intersections, and stabilizes training in deeper network architectures.
By combining multiple heads, the mechanism approximates an ensemble, increasing parameterization capacity and enabling robust modeling of diverse traffic dynamics to capture the joint interactions among heterogeneous entities such as vehicles and environmental constraints for scenario synthesis, as shown in~\cref{eq:multihead}. 

\begin{equation}
\label{eq:multihead}
\begin{aligned}
\text{MultiHead}(Q, K, V) &= \text{Concat}(\text{head}_1, \ldots, \text{head}_h)W^0 \\
\text{where} \quad \text{head}_i &= \text{Attention}(QW_i^Q, KW_i^K, VW_i^V)
\end{aligned}
\end{equation}
To efficiently integrate the reward, \tool  employs a proxy model consisting of a multi-layer neural network trained on the weighted action sequence set \( S_a \).
The labels are the scenario-specific reward values \( R_s \) computed from traces.
Without this proxy model, candidate reward values during training would require simulator evaluations, which would significantly increase computational time.
The proxy model approximates the reward as defined in~\cref{eq:reward_computation} to effectively weight high-risk violations.
The \tool generative model is trained to assign generation probabilities to sequences such that they correlate with their associated rule violation scores.
This training process minimizes a loss function for a batch of generated sequences as presented in~\cref{eq:loss}.
\begin{equation}
\label{eq:loss}
L(\psi) = \frac{1}{B} \sum_{b=1}^B \left( f(\log P_{\psi}(A_s), \log Z) - g(\log R(A_s)) \right)^2
\end{equation}
where $\psi$ represents the model's parameters, $P_{\psi}(A_s)$ is the generation probability for sequence $A_s$, $R(A_s)$ is its rule violation score, $B$ is the batch size, $f$ combines the model’s $P_{\psi}(A_s)$ with $\log Z$ to form the log of the “flow” for sequence $A_s$, $g$ filters out invalid sequences, ensuring that the severity of rule violations directly influences the loss, and $Z$ normalizes the model’s probability distribution over all possible action sequences to match reward $R(A_s)$.
By minimizing this loss, the model's parameters are adjusted to make the generation probability of a sequence proportional to its weighted reward.  The legal weighting mechanism is crucial in this process, as it amplifies the reward for severe violations, which provides a stronger learning signal.
This provides a strong directional signal, guiding the model to focus its capacity on the rare and complex features characteristic of critical-yet-hard-to-find scenarios, rather than treating all violations equally.
This mechanism ensures that \tool generates high-reward, safety-critical scenarios for testing.

\subsubsection{Reward-Driven Action Sequences Crafting}
In this phase, \tool generates traffic scenarios by sampling action sequences based on weighted laws and interactions.
This process aims to construct action sequences that maximize the expected reward, which reflects how strongly the scenarios trigger traffic law violations.
The reward function, introduced in~\cref{eq:reward_computation}, quantifies the severity and frequency of violations, guiding the sampling process toward high-risk, safety-critical scenarios valuable for ADS.
The scenario generation process is formalized as a state transition chain \( \Gamma \), as defined in~\cref{eq:chain}.
\begin{equation}
\label{eq:chain}
\Gamma = \langle s_0 \xrightarrow{a_0} s_1 \xrightarrow{a_1} \cdots \xrightarrow{a_t} s_t \longrightarrow s_f\rangle
\end{equation}
where $s_t$ denotes the system state at step $t$, encapsulating parametric configurations such as time, weather conditions, and vehicles, and other relevant parameters; $a_t$ represents the action executed at $s_t$, modifying scenario parameters through operations like setting weather to rainy or adjusting the speed of NPC vehicles; $s_0$ is the initial empty configuration; and $s_f$ is the terminal state representing a syntactically valid and executable test scenario.
At each step, the model selects the next action based on a reward-driven policy that biases the sampling toward high-reward behaviors.

\tool is trained using \( R \) to favor sequences with greater violations by minimizing the loss function, as shown in~\cref{eq:loss}, over a batch of generated sequences.
This process tunes the multi-head attention to focus on scenario features and interactions that drive higher rule violation scores, increasing the likelihood of generating such sequences.

Once the action sequence is crafted, it is decoded into an executable scenario script format.
The decoding process transforms the abstract sequence into concrete simulation parameters, e.g., time settings, vehicle speeds, signal states, and environmental conditions.
This ensures that the generated scenarios can be reproduced and tested in a simulation environment.

\subsection{Scenario Testing}
As described in~\cref{alg:scenario-testing}, scenario testing evaluates the behavior of the ADS under various traffic conditions with respect to the specified traffic laws.
For each scenario generated in the previous scenario generation phase, \tool executes the ADS within the scenario, monitors its behavior, and records the resulting execution trace \(\pi\) (\cref{line:run_avs}). This trace is then analyzed against the relevant traffic law clauses to identify any violations caused by the AV's actions (\cref{line:foreach_formula}–\cref{line:remove_formula}). All identified violations are stored in \(Violations\) and reported to the user upon completion of the testing process.

\begin{algorithm}
\caption{Scenario Testing}
\label{alg:scenario-testing}
\KwIn{$B$: Set of generated testing scenarios\\
      $\phi(\Theta)$: Set of traffic law formulas\\
      ADS: Automated Driving System}
\KwOut{Violations: List of violated formulas and corresponding scenarios}

\SetKwFunction{FMain}{ScenarioTesting}
\SetKwProg{Fn}{Function}{:}{}
\SetKwInOut{Input}{Input}\SetKwInOut{Output}{Output}

\setcounter{AlgoLine}{0}
\Fn{\FMain{$B$, $\phi(\Theta)$, ADS}}{
    $Violations \gets \emptyset$ \label{line:init_violations}\;
    \ForEach{scenario $b \in B$}{
        \label{line:foreach_scenario}
        Run $ADS$ on $b$ to generate trace $\pi$\;
        \label{line:run_avs}
        \ForEach{formula $\phi_i \in \phi(\Theta)$}{
            \label{line:foreach_formula}
            Compute robustness $r_{\text{STL}}(\phi_i, \pi, t)$\;
            \label{line:compute_robustness}
            \If{$r_{\text{STL}}(\phi_i, \pi, t) < 0$}{
                Add $(\phi_i, b)$ to $Violations$\;
                \label{line:add_violation}
                Remove $\phi_i$ from $\phi(\Theta)$
                \label{line:remove_formula}
            }
        }
    }
    \Return $Violations$\;
    \label{line:return_violations}
}
\end{algorithm}

When analyzing the recorded trace \(\pi\), \tool employs RTAMT~\cite{nivckovic2020rtamt}, an STL verification library, to compute the robustness of the trace with respect to each traffic law specification.
Key vehicle state information, such as speed, acceleration, and relative position, is extracted from \(\pi\) and combined with contextual data about the road and surrounding agents.
RTAMT then computes the robustness value \( r_{\text{STL}}(\phi_i, \pi, t) \), where a positive value indicates that the trace \(\pi\) satisfies the law \(\phi_i\), and a negative value signifies a violation.
The magnitude of the robustness reflects the degree of satisfaction or severity of the violation.
Violation detection in \tool is determined by the deterministic semantics of STL, where robustness values quantify the exact degree of law compliance beyond a simple binary result.
If a violation is detected, the algorithm records the offending law \(\phi_i\) and the corresponding scenario \(b\) in the violation list.
\tool then removes \(\phi_i\) from the traffic law set \(\phi(\Theta)\) to prevent redundant verification in subsequent scenarios.


\section{Evaluation}\label{sec:Evaluation}
In this section, we present the evaluation results with respect to \tool's effectiveness and efficiency. We begin by describing the experimental setup, the applied benchmark datasets, and the two state-of-the-art tools for comparison.
We then address four research questions, detailing the corresponding experimental results and analyses.

\subsection{Experimental Setup}
We have compared \tool against two state-of-the-art ADS testing frameworks, ABLE~\cite{zhang2023testing} and LawBreaker~\cite{sun2022lawbreaker}, through a group of experiments.
These baselines are selected based on three criteria: (1) they target complete ADS testing, (2) they support concrete scenario generation within a simulator, and (3) they enable traffic law compliance detection.
The target of our evaluation is Baidu Apollo v8.0~\cite{apollo_v8}, an industrial-grade open-source ADS. For all experiments, the Apollo system was integrated and executed within the CARLA simulator~\cite{dosovitskiy2017carla}, which provided the traffic environment.
Each scenario simulation terminates when the ego vehicle reaches its destination, is involved in a collision, or times out due to blockage.
To ensure a fair comparison, the baseline tools, ABLE and LawBreaker, were also configured to test the same Apollo instance within CARLA.
Since both ABLE and LawBreaker were originally implemented on the LGSVL~\cite{rong2020lgsvl} simulator, we ported their implementations to CARLA by adapting perception and control interfaces while preserving their core algorithms. This ensured their functionalities were retained in the CARLA environment.
\tool was trained on a dataset of 12,800 scenarios generated by LawBreaker, alongside a set of 81 STL-based traffic law violation formulas. These formulas were derived from a subset of Chinese traffic regulations that could be directly translated into STL expressions.
The selection of these regulations was based on their ability to be formalized in STL, focusing on the most common and significant traffic violations.
Our implementation employs China's ``Administrative Measures for Road Traffic Safety Violation Scoring''~\cite{Ministry_of_Public_Security_2021} and ``Motor Vehicle Driver Training Syllabus''~\cite{MOT_MPS_2022} as instances of these sources, which further support the risk weighting of violations.
While the set covers a broad range of key traffic laws, it does not represent the entire spectrum of Chinese traffic laws.
The model was trained for 1000 epochs to ensure convergence.
For both ABLE and LawBreaker, we obtained the latest versions from their authors and initialized them using an identical set of Chinese traffic laws.

In total, we conducted four comparative experiments to answer the research questions:
\begin{itemize}
    \item RQ1: How effective is \tool in detecting traffic law violations?
    \item RQ2: How does \tool perform with respect to the diversity of generated scenarios?
    \item RQ3: How efficient is \tool in terms of computational resources and runtime?
    \item RQ4: How do individual components impact \tool's overall performance?
\end{itemize}

The first three research questions focus on comparing \tool with other state-of-the-art tools in terms of effectiveness, scenario diversity, and efficiency.
The fourth question examines the contribution of individual components and the overall robustness of \tool. 
Furthermore, we investigate the reliability of the weighting module across various LLM backends while comparing the proxy model with the simulator to assess reward approximation fidelity and efficiency.
To address this, we conducted a series of ablation studies.
All experiments were conducted on a machine equipped with a 32-core Intel Core i9-13900K CPU, an NVIDIA RTX 4090 GPU, and 32 GB of system memory.
To mitigate randomness, all experiments were conducted three times, and the reported results are averaged over these runs.
The core of \tool's scenario generation model is a Transformer encoder configured with 16 layers, a hidden dimension of 1024, and 8 attention heads.

The experimental results are categorized based on different road structure configurations: a T-shaped intersection (S1), an intersection with single-direction roads (S2), an intersection with bidirectional roads (S3), and an intersection combining a single-direction road and a bidirectional road (S4).
This classification is consistent with that used in prior work~\cite{avunit2021SpecificationbasedAutonomous}.
These road structures were selected for their ability to broadly represent the majority of real-world driving scenarios.
We evaluated \tool using 1024 scenarios for S1 and S2, and 512 scenarios for S3 and S4.

\begin{figure*}[!ht]
    \centering
    \includegraphics[width=0.8\textwidth]{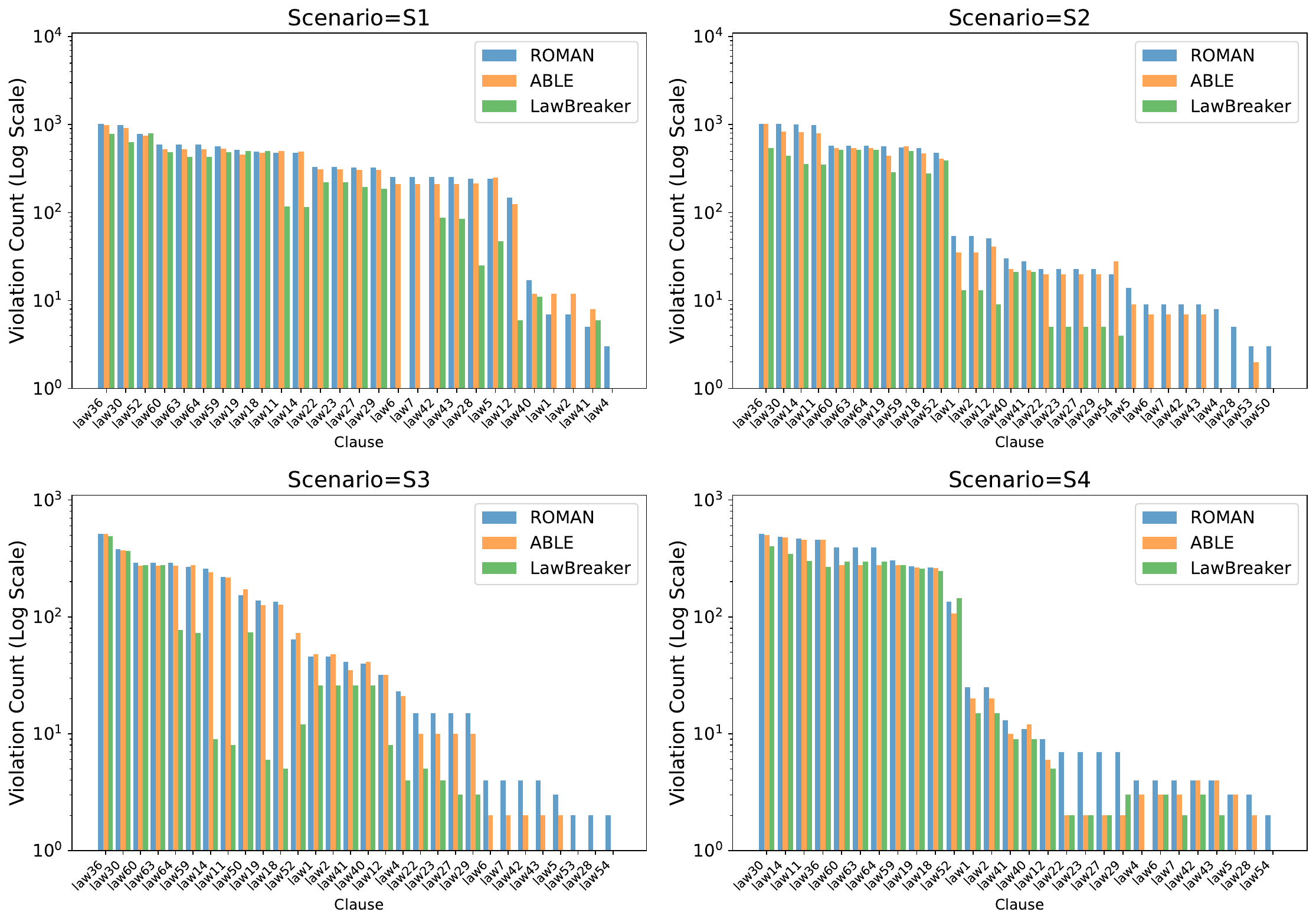} 
    \caption{Comparison of Violation Counts in Scenarios S1 to S4}
    \label{fig:law}
\end{figure*}

\subsection{RQ1: How effective is \tool in detecting traffic law violations?}

To address this question, we executed all three tools and compared them based on the number and types of detected traffic law violations.

As shown in~\cref{fig:law}, we present the violation counts for all three tools, with each subfigure depicting the results for a specific kind of road structure configuration using a logarithmic scale.
Certain laws (e.g., \texttt{law63} and \texttt{law52}) may be violated across multiple road structures, hence their repeated appearance in multiple subfigures.
From S1 to S4, \tool consistently generated more violation scenarios on the majority of traffic law clauses.
\tool successfully consistently provided broader coverage of traffic law clauses than ABLE and LawBreaker across the four types of road structures.
In contrast, ABLE failed to cover 1 clauses under S1, 3 clauses under S2, and 3 clauses under S3.
LawBreaker showed even lower coverage, missing 5, 9, 8, and 4 clauses under S1, S2, S3, and S4, respectively.
Notably, for certain clauses and road structures, only \tool was able to produce scenarios that triggered violations, for instance, \texttt{law4} and \texttt{law50} under the single-direction road intersection (S2), and \texttt{law53}, \texttt{law28}, and \texttt{law54} under the bidirectional road intersection (S3).
We provide a detailed explanation in~\cref{tab:violations} for all traffic laws that only \tool generated violation scenarios.
Most of these laws stipulate when a vehicle is permitted to move; failing to test them on an ADS may conceal significant safety deficiencies and potentially result in fatal traffic incidents.


These results highlight that \tool substantially outperforms the other two tools in terms of violation coverage across diverse traffic scenarios.
\begin{table*}
\centering
\caption{Traffic Law Details that Only \tool Generated Violation Scenarios}
\label{tab:violations}
\begin{tabular}{lll}
\toprule
\textbf{Law ID in Evaluation} & \textbf{Chinese Traffic Law Article} & \textbf{Traffic Law Detail} \\
\midrule
Law4  & Article 38, Sub-item 2 & Vehicles past the stop line may proceed on yellow. \\
\midrule
Law28 & Article 47 & \makecell[l]{Before overtaking, a vehicle must keep a safe distance, pass on the left, \\then signal right and return to its lane.} \\
\midrule
Law50  & Article 52, Sub-item 2-4 & \makecell[l] {At unmarked intersections, turning vehicles yield to through vehicles, \\ and right turns yield to left turns.} \\
\midrule
Law53 & Article 53 & In intersection congestion, vehicles must wait outside and not enter. \\
\midrule
Law54 & Article 53 & \makecell[l]{When traffic is slow or queued, vehicles must line up in order, \\ not weave, overtake, or stop on crosswalks or grid lines.} \\
\bottomrule
\end{tabular}
\end{table*}


To further assess \tool's effectiveness, \cref{tab:violation_distribution} presents five fine-grained metrics that offer deeper insight into the identified violations.
The column ``Mean'' represents the arithmetic average number of violations per scenario, and the column ``Max'' indicates the maximum number of violations observed in a single scenario.
The next three columns report the number of scenarios in which law violations exceeds predefined thresholds, with a higher threshold indicating greater safety risk.
We define three thresholds here: ``High Rate'' $>$ 6, ``High Rate'' $>$ 8, and ``High Rate'' $>$ 10.
\tool significantly outperformed both ABLE and LawBreaker in generating violation scenarios across all four types of road structures (S1–S4), excelling in key evaluation metrics: Mean, Max, and High Rate.
To illustrate this, we highlight results from S3 as a representative case.

\tool achieved an average violation count that exceeded ABLE's by 7.91\% and LawBreaker's by 55.96\%.
For the ``Max'' metric, \tool produced a scenario containing 18 traffic law violations, whereas ABLE and LawBreaker reached only 16 and 9 violations, respectively.
In terms of ``High Rate metrics'', representing the proportion of scenarios exceeding predefined violation thresholds, \tool consistently outperformed the baselines.
In particular, for the ``High Rate >8'' metric, which reflects the share of high-risk scenarios, \tool generated 24\% of such cases, which is 26.3\% more than ABLE’s 19\% and 71.4\% more than LawBreaker’s 14\%. For the ``High Rate $>10$'' metric, \tool produced 9\%, which is 80\% more than ABLE’s 5\% and significantly more than LawBreaker’s 0\%.
This superior performance in the complex, bidirectional traffic of the S3 scenario stems from the ability of \tool's multi-head attention mechanism to simultaneously model the spatiotemporal interactions between multiple NPC vehicles during critical maneuvers.

\begin{table}[!ht]
    \centering\tabcolsep=1pt
    \caption{Violation Count Distribution. The best results are highlighted with a wavy underline.}
    \label{tab:violation_distribution}
    \footnotesize 
\resizebox{\linewidth}{!}{
    \begin{tabular}{clccccc}
        \toprule
       Scenario & Approach & Mean &Max & High  Rate  $>$6 &High  Rate  $>$8 &High  Rate $>$10 \\
        \midrule
        \multicolumn{1}{c}{\multirow{3}{*}{\normalsize S1}} & \tool & \uwave{\textbf{9.86}} & \uwave{\textbf{22}} & \uwave{\textbf{75}}\%& \uwave{\textbf{61}}\% & \uwave{\textbf{42}}\%\\
           & ABLE & 9.12 & 20 & 64\% & 49\% & 38\%\\
           & LawBreaker & 4.90 & 11 & 40\% & 14\% & 6\%\\
        \midrule
        \multicolumn{1}{c}{\multirow{3}{*}{\normalsize S2}} & \tool & \uwave{\textbf{8.07}} & \uwave{\textbf{19}}  & \uwave{\textbf{70}\%} & \uwave{\textbf{46}\%} & \uwave{\textbf{17}\%} \\
           & ABLE & 7.23 & 17 & 58\%& 38\% & 15\% \\
           & LawBreaker & 3.44 & 12 &  31\%&  10\% & 2\%\\
        \midrule
        \multicolumn{1}{c}{\multirow{3}{*}{\normalsize S3}} & \tool & \uwave{\textbf{6.41}} & \uwave{\textbf{18}} &  \uwave{\textbf{43}\%}&  \uwave{\textbf{24}\%} &  \uwave{\textbf{9}\%}\\
           & ABLE & 5.94 & 16 &  39\%&  19\% &  5\%\\
           & LawBreaker & 4.11 & 9 &  22\%&  14\% &  0\% \\
        \midrule
        \multicolumn{1}{c}{\multirow{3}{*}{\normalsize S4}} & \tool & \uwave{\textbf{7.15}} & \uwave{\textbf{15}}  & \uwave{\textbf{57\%}} & \uwave{\textbf{31\%}} & \uwave{\textbf{5\%}}\\
           & ABLE & 7.07 & 13  & 51\%& 29\% & 4\% \\
           & LawBreaker & 6.78 & 11 & 46\% & 21\% & 2\% \\
        \bottomrule
    \end{tabular}}
\end{table}

\begin{figure*}[!ht]
    \centering
    \includegraphics[width=0.7\textwidth]{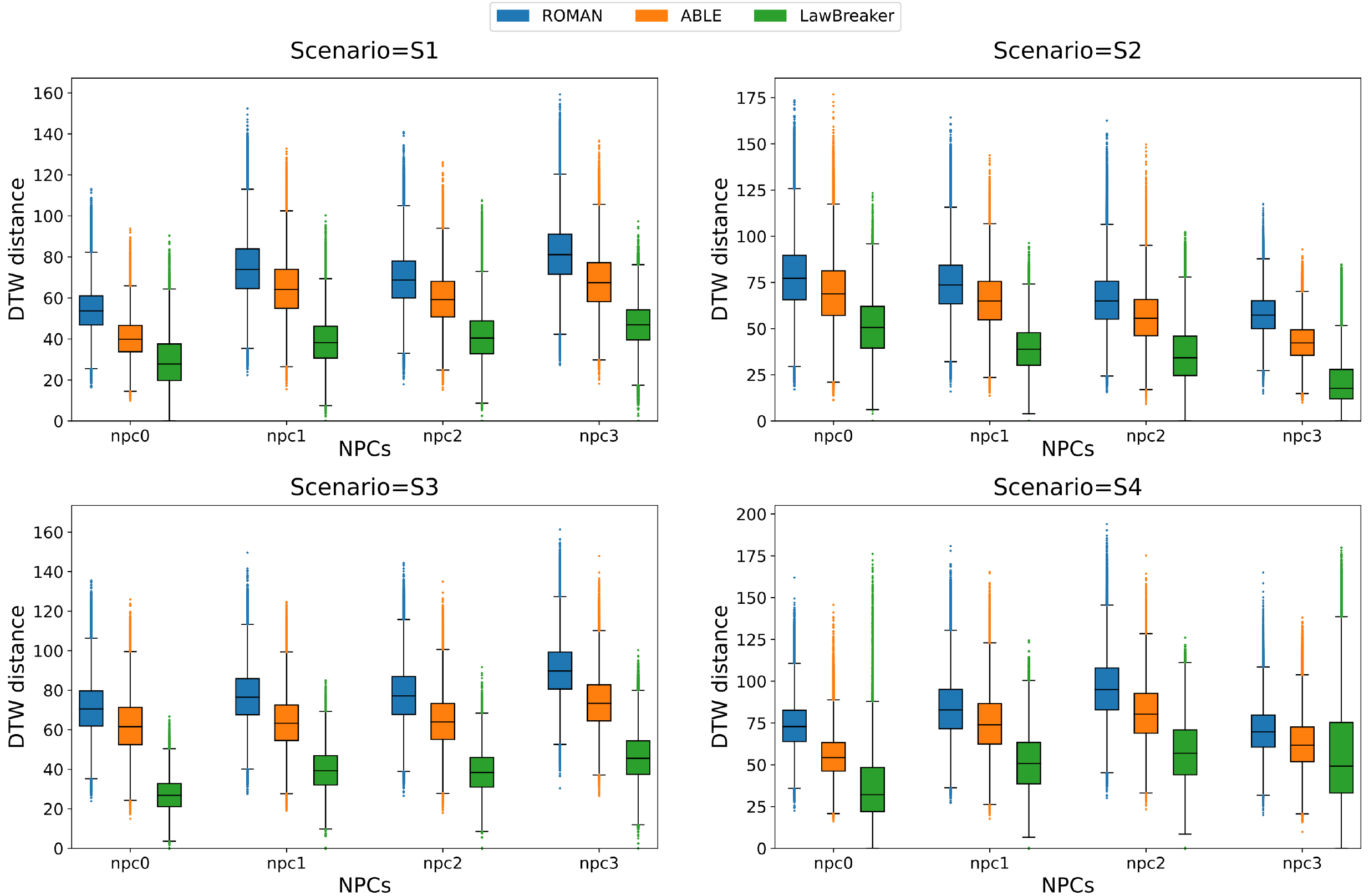} 
    \caption{Scenario Diversity Comparison Using DTW Distance Across S1-S4}
    \label{fig:rq2}
\end{figure*}

\begin{mdframed}
\textbf{Takeaways:}  \tool outperformed LawBreaker and ABLE in generating high-quality violation scenarios with broader law coverage.
\end{mdframed}

\subsection{RQ2: How does \tool perform with respect to the diversity of generated scenarios?}

To answer this question, we evaluated the effectiveness of \tool on scenario diversity.
Complex traffic scenarios enhance ADS testing by exposing systems to varied traffic conditions, improving the detection of challenging situations.
A larger DTW~\cite{muller2007dynamic} distance indicates greater diversity.
For two trajectories $ X = \{x_1, x_2, \ldots, x_n\} $ and $x' = \{x_1', x_2', \ldots, x_m'\} $, where $ x_i $ and $ x_j' $ are 2D coordinates of an NPC's positions, the DTW distance is defined as~\cref{eq:dtw}.
\begin{equation}
\label{eq:dtw}
\text{DTW}_q(x, x') = \min_{\pi \in \mathcal{A}(x, x')} \left( \sum_{(i,j) \in \pi} d(x_i, x_j')^q \right)^{\frac{1}{q}}
\end{equation}
where $ q \geq 1 $ is a parameter that adjusts the sensitivity of the metric to point-wise distances, with larger $ q $ emphasizing significant deviations between trajectories, the term $ d(x_i, x_j') $ is the Euclidean distance between points $  x_i $  and $  x_j' $ , and the minimization is over all admissible alignment paths $ \pi \in \mathcal{A}(x, x')$, an admissible path \(\pi\) is a sequence of index pairs \((i_k, j_k)\) that aligns the trajectories while satisfying monotonicity constraints~\cite{tavenard.blog.dtw}.

As shown in~\cref{fig:rq2}, we present the DTW distance distributions for each NPC across \tool, ABLE, and LawBreaker, under the four road structure categories. To enable fair comparison, we fixed the number of NPCs in each generated scenario.
For each pair of scenarios within the same road structure category, we computed the DTW distance and finally visualized the resulting distributions using box plots.
For all NPCs except \texttt{npc3} under S4, \tool exhibited higher values across \textbf{all} percentiles in the DTW distance distributions.
Even for \texttt{npc3} under S4, only LawBreaker surpassed \tool at the 100th percentile, while \tool maintained higher values at the remaining four key percentiles.
These results demonstrate that \tool consistently outperforms the other two tools in generating more diverse scenarios.

\tool achieved an average median DTW distance of 74.05 across all scenarios and NPCs, exceeding ABLE's 62.19 by 19.07\% and LawBreaker's 39.62 by 86.90\%, for the 100th percentile, representing greater trajectory variability, \tool averaged 113.40, whereas ABLE reached 100.71 and LawBreaker only 80.56, In terms of 0th percentile, reflecting baseline diversity, \tool averaged 35.35, surpassing ABLE's 24.36 and LawBreaker's 5.80.
At the 25th percentile, \tool achieved an average of 64.61, exceeding ABLE's 52.97 and LawBreaker's 30.52, similarly, at the 75th percentile, \tool averaged 84.13, surpassing ABLE's 72.06 and LawBreaker's 50.54.
Critically, we observed LawBreaker's 0th percentile values dropping to 0 for \texttt{npc0} under S1, \texttt{npc2}, \texttt{npc3} under S2, and \texttt{npc0}, \texttt{npc3} under S4, which we attribute to premature convergence in its genetic algorithm design, causing the generation of numerous identical trajectories for these specific NPCs.
In particular, in Scenario S4 for \texttt{npc2}, \tool's median hit 95.00, outperforming ABLE's median 80.36 by 18.21\% and LawBreaker's 56.89 by 66.98\%, demonstrating superior trajectory diversity due to its multi-head attention mechanism capturing complex traffic interactions.

\tool's outstanding effects in diversity stem from its multi-head attention mechanism, which captures the subtle correlations across multiple features, enabling varied trajectory modeling.
In contrast, ABLE's distributions are much narrower, with medians rarely exceeding 80, indicating limited variability.
LawBreaker's distributions are even more constrained, with medians typically below 60 and 0th percentile often near or below 10, along with fewer outliers, reflecting reduced diversity in complex scenarios.
In summary, \tool's use of a multi-head attention mechanism facilitates the generation of diverse scenarios, thereby enhancing ADS testing by covering a broader range of traffic dynamics.
\begin{mdframed}[style=mystyle]
\textbf{Takeaways:} \tool's multi-head attention increases scenario diversity, with DTW distributions outperformed ABLE and LawBreaker.
\end{mdframed}

\subsection{RQ3: How efficient is \tool in terms of computational resources and runtime?}
To answer this question, we evaluated the efficiency of all three tools by measuring their training time and inference time for scenario generation across different road structure categories.
We focused on both training and inference time for scenario generation.
\begin{table}[H]
    \centering\tabcolsep=2pt
    \caption{Training and Inference Time of Different Approaches}
    \label{tab:execution_time}
    \normalsize  
    \renewcommand{\arraystretch}{1} 
\resizebox{\linewidth}{!}{
    \begin{tabular}{lrrrrrrr}
        \toprule
        \multirow{2}{*}{Scenario} &
        \multicolumn{2}{c}{\tool} &
        \multicolumn{2}{c}{ABLE} &
        \multicolumn{2}{c}{LawBreaker} \\
        \cmidrule(lr){2-3} \cmidrule(lr){4-5} \cmidrule(lr){6-7}
         & Training & Inference & Training & Inference & Training & Inference \\
        \midrule
        S1 & 9011s  & 44s & 6504s  & 33s & -- & 37s \\
        S2 & 8456s  & 47s & 6054s  & 36s & -- & 52s \\
        S3 & 14813s & 52s & 10224s & 38s & -- & 62s \\
        S4 & 13664s & 53s & 11664s & 41s & -- & 46s \\
        \bottomrule
    \end{tabular}}
\end{table}

As shown in~\cref{tab:execution_time}, \tool required substantially longer training time than ABLE and LawBreaker, while the inference time across the three tools remained relatively close.
The longer runtime of \tool is primarily due to its extended training phase, resulting from its computationally intensive multi-head attention mechanism.
This trade-off focuses on generating higher-quality, more diverse scenarios, which is essential for robust autonomous driving validation.
However, despite the longer training time, \tool's inference time is similar to ABLE's, ranging from 44s to 53s, while ABLE's inference time ranges from 33s to 41s.
LawBreaker employs a genetic algorithm to generate traffic law violation scenarios.
Consequently, it does not require a training phase and only incurs inference time.
While it demonstrates notable efficiency in inference, its overall quality is substantially lower than that of the other two tools, failing to cover a significant number of traffic law clauses and producing the least diverse set of scenarios.
For ABLE, its underlying GFlowNet model leads to shorter training and inference time compared to \tool’s multi-head attention network. However, \tool still demonstrates comparable efficiency.
With an additional 0.57 to 1.28 hours of total training and inference time, \tool produces a higher-quality set of scenarios, covering all traffic law clauses and consistently exhibiting greater diversity.
Moreover, even for the most time-consuming road structure category S3, \tool completed both training in 14,813s and inference in 52s within a total time of 4.13 hours, which remains practical for ADS testing.

\begin{mdframed}[style=mystyle]
\textbf{Takeaways:} \tool excels at generating high-quality, diverse scenarios for ADS testing, with training time longer but inference time comparable to ABLE, ensuring a balanced computational cost.
\end{mdframed}

\subsection{RQ4: How do individual components impact \tool's overall performance?}
To answer this question, we conducted a group of ablation studies to evaluate the effect of two key components on \tool's performance.
First, we assessed the impact of the traffic law weighting mechanism by comparing \tool with weighting and \tool without weighting (\tool-unweighting).
Second, we evaluated the effect of the number of attention heads by comparing \tool with eight attention heads, four attention heads, and two attention heads.
Experiments with more than eight heads were omitted due to GPU memory constraints.
Third, we evaluated the impact of training and generation temperatures on model performance and scenario diversity.

\begin{figure*}[!ht]
    \centering
        \centering
        \includegraphics[width=0.98\linewidth]{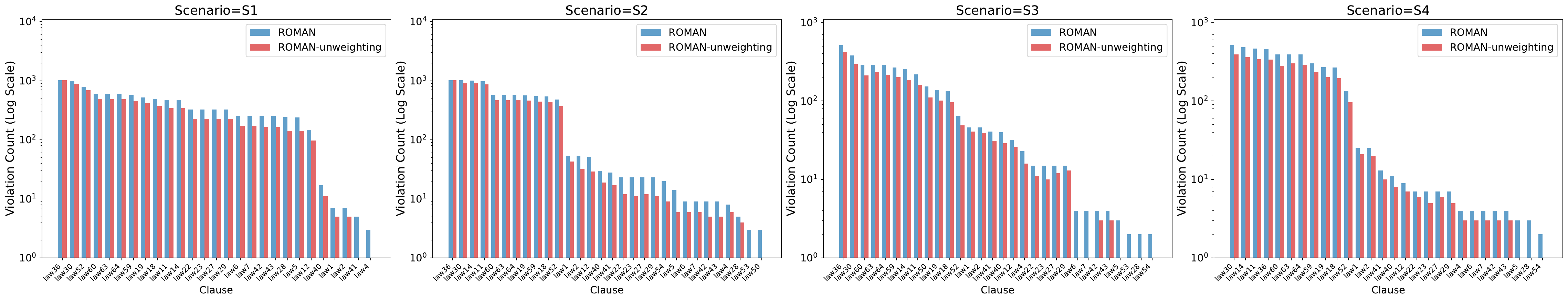}
        \caption{Violation Count Distribution with Weighting and Unweighting}
        \label{fig:RQ3law}
        \includegraphics[width=0.98\linewidth]{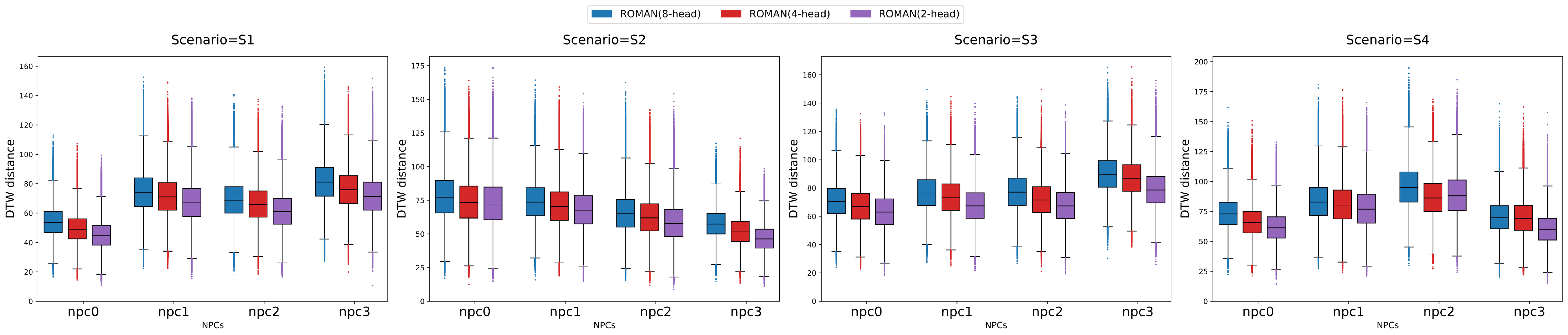}
        \caption{DTW Distance under Different Attention}
        \label{fig:rq32}
\end{figure*}

~\cref{fig:RQ3law} shows the violation counts for various traffic law clauses.
Taking Scenario S3 as a representative example, \tool successfully triggered violations for critical high-risk clauses such as \texttt{Law6}, \texttt{Law7},  \texttt{Law5}, \texttt{Law53}, \texttt{Law28}, and \texttt{Law54}.
In contrast, \tool-unweighting showed reduced coverage for these prioritized clauses.
This pattern of \tool achieving broader coverage of high-weight, safety-critical laws persisted consistently in Scenarios S1, S2, and S4.
This trend persists in S1, S2, and S4, with \tool demonstrating broader coverage of critical clauses such as Law28 and Law4, which are weighted by the weighting mechanism.
This suggests that the traffic law weighting mechanism enhances the focus on critical safety-related laws, improving overall law coverage in the generated scenarios.

~\cref{fig:rq32} presents the scenario diversity through DTW distances.
\tool with 8 heads consistently demonstrated superior trajectory diversity, achieving the highest average median DTW distance of 74.05 compared to 69.92 for 4-head and 65.64 for 2-head.
It also achieved the highest average 100th percentile of 113.40 versus 108.82 for 4-head and 104.28 for 2-head, along with the highest average 0th percentile of 35.35 against 31.65 for 4-head and 27.61 for 2-head.
Additionally, it recorded the highest average 25th percentile of 64.61, surpassing 60.59 for 4-head and 56.36 for 2-head. Likewise, at the 75th percentile, \tool achieved the highest average of 84.13, outperforming 79.89 for 4-head and 75.53 for 2-head.
\tool with 8 heads demonstrated superior trajectory diversity, achieving an average median DTW distance of 5.90\% higher than 4-head configurations and 12.81\% higher than 2-head configurations.
This performance advantage confirms its enhanced capacity for modeling complex traffic interactions, with the 8-head configuration showing the most significant improvement across all evaluation metrics compared to fewer-head setups.

This pattern demonstrates that a higher number of attention heads will enhance the modeling of complex interactions, resulting in more diverse scenarios.
These findings highlight that increasing the number of attention heads to 8 significantly boosts scenario diversity in ADS testing, as evidenced by the consistently higher median DTW distances and broader distributions across all NPCs and scenarios.

\begin{table*}[!ht]
\centering\tabcolsep=4pt
\caption{Evaluation of Training and Generation Temperatures. For each metric, `↑` indicates higher is better, and `↓` indicates lower is better. The optimal configuration (100, 1.00) is highlighted with a wavy underline.}
\label{tab:temp_ablation}
\begin{tabular}{cc|cccccccc}
\toprule
\textbf{Train Temp.} & \textbf{Gen Temp.} & \textbf{Completeness ↑} & \textbf{Satisfaction ↑} & \textbf{Distinct-1 ↑} & \textbf{Distinct-2 ↑} & \textbf{Distinct-3 ↑} & \textbf{Self-BLEU ↓} & \textbf{Entropy ↑} & \textbf{Coverage ↑} \\
\midrule
\multirow{3}{*}{75} & 0.25 & 0.96 & 0.95 & 0.08 & 0.15 & 0.21 & 0.65 & 5.50 & 0.45 \\
& 0.50 & 0.95 & 0.93 & 0.11 & 0.20 & 0.28 & 0.64 & 5.40 & 0.49 \\
                      & 0.75 & 0.94 & 0.93 & 0.16 & 0.30 & 0.40 & 0.48 & 6.20 & 0.60 \\
                      & 1.00 & 0.92 & 0.91 & 0.20 & 0.42 & 0.55 & 0.30 & 6.80 & 0.82 \\
                      & 1.25 & 0.75 & 0.68 & 0.24 & 0.56 & 0.69 & 0.19 & 7.20 & 0.90 \\
\midrule
\multirow{3}{*}{100} & 0.25 & 0.95 & 0.94 & 0.09 & 0.18 & 0.25 & 0.68 & 5.70 & 0.48 \\
& 0.50 & 0.95 & 0.92 & 0.13 & 0.22 & 0.30 & 0.60 & 6.00 & 0.55 \\
                           & 0.75 & 0.93 & 0.92 & 0.17 & 0.33 & 0.43 & 0.45 & 6.40 & 0.65 \\
                            & \uwave{\textbf{1.00}} & \uwave{\textbf{0.91}} & \uwave{\textbf{0.90}} & \uwave{\textbf{0.25}} & \uwave{\textbf{0.48}} & \uwave{\textbf{0.61}} & \uwave{\textbf{0.25}} & \uwave{\textbf{7.04}} & \uwave{\textbf{0.88}} \\
                           & 1.25 & 0.74 & 0.65 & 0.28 & 0.60 & 0.73 & 0.15 & 7.40 & 0.94 \\
\midrule
\multirow{3}{*}{125} & 0.25 & 0.89 & 0.88 & 0.09 & 0.19 & 0.26 & 0.66 & 5.72 & 0.49 \\
& 0.50 & 0.87 & 0.88 & 0.14 & 0.28 & 0.34 & 0.52 & 6.04 & 0.59 \\
                           & 0.75 & 0.87 & 0.86 & 0.18 & 0.34 & 0.44 & 0.45 & 6.41 & 0.63 \\
                           & 1.00 & 0.86 & 0.84 & 0.25 & 0.49 & 0.62 & 0.24 & 7.06 & 0.88 \\
                           & 1.25 & 0.70 & 0.61 & 0.30 & 0.64 & 0.77 & 0.14 & 7.42 & 0.95 \\
\bottomrule
\end{tabular}
\end{table*}

~\cref{tab:temp_ablation} presents the impact of training temperatures 75, 100, and 125 and generation temperatures 0.25, 0.50, 0.75, 1.00, and 1.25 on scenario quality. The quality of generated scenarios is measured by the following metrics for validity and diversity.

Validity Metrics: Assess structural and logical correctness.
\begin{itemize}
    \item Completeness: Verifies the presence of all required fields for syntactic correctness.
    \item Satisfaction: Verifies adherence to logical constraints (e.g., parameter ranges).
\end{itemize}

Diversity Metrics: Measure lexical and sequential diversity.
\begin{itemize}
    \item Distinct-1, -2, -3~\cite{li2015diversity}: Measure token diversity by calculating the proportion of unique single words, word pairs, and word triplets among all generated words, pairs, and triplets.
    \item Self-BLEU~\cite{papineni2002bleu}: Measures similarity via the average pairwise similarity between generated sequences, where lower values indicate higher diversity.
    \item Entropy~\cite{shannon1948mathematical}:
    Measures the unpredictability of the token distribution.
    \item Coverage: Calculates the percentage of a core vocabulary used in the output.
\end{itemize}

Our analysis shows that training temperature governs what the model can learn, while generation temperature governs how that capacity is expressed. We found an opposing trend between validity and diversity, where higher generation temperatures improve diversity at the cost of validity. For instance, at a training temperature of 75, increasing the generation temperature from 0.25 to 1.25 caused Completeness to fall from 0.96 to 0.75 and Satisfaction to drop from 0.95 to 0.68.
In parallel, diversity metrics improved over a similar range, with Distinct-1 rising from 0.08 to 0.20 and Self-BLEU dropping from 0.65 to 0.30.
The best balance was achieved with a training temperature of 100 and a generation temperature of 1.00.
This setting maintained high validity, with Completeness at 0.91 and Satisfaction at 0.90, while also delivering strong diversity. Specifically, the Distinct-1, Distinct-2, and Distinct-3 scores rose to 0.25, 0.48, and 0.61, and Self-BLEU declined to 0.25.
In contrast, increasing the training temperature to 125 incurred a notable validity cost, with Completeness and  Satisfaction falling to 0.70 and 0.61, despite producing higher diversity.
Therefore, the 100/1.00 pairing offers the most balanced setting for generating scenarios that are both structurally sound and richly varied.

\begin{table}[t]
\centering
\caption{Cross-Model Weight Consistency Analysis}
\label{tab:model_consistency}
\footnotesize
\setlength{\tabcolsep}{6pt}
\begin{tabular}{l cccc cc}
\toprule
Model & \multicolumn{4}{c}{Spearman Correlation} & MAE & Avg. Tokens \\
\cmidrule(r){2-5}
 & GPT & Gemini & Claude & Qwen & & \\
\midrule
GPT    & 1.000 & 0.948 & 0.906 & 0.942 & 0.039 & 5984 \\
Gemini & 0.948 & 1.000 & 0.925 & 0.954 & 0.031 & 6213 \\
Claude & 0.906 & 0.925 & 1.000 & 0.921 & 0.068 & 5418 \\
Qwen   & 0.942 & 0.954 & 0.921 & 1.000 & 0.035 & 5166 \\
\bottomrule
\end{tabular}
\end{table}
\cref{tab:model_consistency} reports the robustness of the weighting protocol across four LLM backends.
All pairwise Spearman correlation coefficients exceed 0.90, indicating a strong and consistent consensus on law-based risk prioritization.
The Mean Absolute Error remains below 0.07 on a 4.0 scale.
In terms of resource cost, the average token consumption per law clause is 6213 (Gemini), 5984 (GPT), 5418 (Claude), and 5166 (Qwen).
Together with expert review by senior traffic police and ADS professionals, these results demonstrate that \tool produces robust, accurate, and model-agnostic risk weights.

\begin{table}[h]
\centering
\caption{Comparison between Proxy Model and Simulator}
\label{tab:proxy_vs_simulator}
\begin{tabular}{@{}lccc@{}}
\toprule
\textbf{Evaluator} & \textbf{Avg. Time} & \textbf{Mean Reward} & \textbf{MRE (\%)} \\ \midrule
Simulator     & 86.8409 s                    & 0.2671              & \multirow{2}{*}{7.35\%} \\
Proxy Model        & 0.0095 s            & 0.2616              &                   \\ \bottomrule
\end{tabular}
\end{table}

\cref{tab:proxy_vs_simulator} evaluates reward approximation using the proxy model.
It achieves a Mean Relative Error (MRE) of 7.35\% with a mean reward of 0.2616, compared with 0.2671 from the simulator, while reducing the per-scenario computational cost from 86.8409s to 0.0095s, corresponding to an improvement of approximately 9000 times.
This efficiency enables extensive iterative training without continuous simulator feedback.

\begin{mdframed}[style=mystyle]
\textbf{Takeaways:} \tool prioritizes regulations with high safety impact, generates diverse scenarios, and achieves robust, efficient, high-fidelity reward approximation through a proxy model.
\end{mdframed}
\section{Threats to Validity}\label{sec:Threats}
First, \tool excels in generating high-quality and complex traffic scenarios, but its computational performance has certain bottlenecks.
The time required for scenario generation by \tool is significantly longer than that of ABLE and LawBreaker, and its computational overhead is relatively high, which may limit its efficiency in large-scale testing tasks. 
This overhead can be mitigated by leveraging multi-GPU parallelism, making \tool more practical and scalable for large-scale scenario generation.

Second, the effectiveness of our LLM-based assessment is closely tied to the quality and completeness of the official documents within our knowledge base. 
While the LLM may inherit certain biases from its training data, expert validation is employed to review and mitigate these biases, improving the overall reliability.

Third, \tool cannot guarantee full coverage of extremely rare edge cases due to constraints in the search space and the stochastic nature of traffic dynamics.
\section{Related Work}\label{sec:Related}
This section reviews scenario generation and diversity methods for ADS, comparing to \tool.

\textbf{Scenario generation} is vital for testing ADS to create diverse scenarios that reveal deficiencies in complex traffic environments. 
LawBreaker~\cite{sun2022lawbreaker} targets traffic law violations for law coverage, while ConfVE~\cite{chen2024misconfiguration} addresses configuration testing framework and TM-fuzzer~\cite{lin2024tm} handles traffic interactions to reveal vulnerabilities. ABLE~\cite{zhang2023testing} uses GFlowNet rewards for better efficiency and diversity but struggles with complex dynamic scenarios. SceGene~\cite{9662987} combines genetic algorithms and natural selection but misses critical high-risk situations. TrafficGAN~\cite{zhang2019trafficgan} generates realistic and diverse traffic scenarios, partially but often focuses on simpler scenarios.
ScenarioFuzz~\cite{wang2024dance} automates script free scenario generation via map crawling and utilizes a GNN based evaluation model to filter high risk seeds.
Recently, VioHawk~\cite{li2024viohawk} focuses on fuzzing hazard zones but introduces NPC vehicles in certain regions to induce violations, along with 42 traffic laws and a modeling format that differ from our STL-based framework. These differences lead to significant performance deviations, making direct comparisons challenging.
\tool generates complex, high-risk scenarios with a traffic law-weighted mechanism and multi-head attention, enhancing ADS test coverage for laws.

\textbf{Scenario Diversity} is vital for ADS testing to reveal system limitations.
Methods like LLM-based generation~\cite{aasi2024generating}, TSC2CARLA~\cite{borchers2025tsc2carla}, clustering-based analysis~\cite{schutt2023clustering}, BehAVExplor~\cite{cheng2023behavexplor}, and CaDRE with deep reinforcement learning~\cite{sun2021corner} generate diverse scenarios but often focus on moderate-risk cases. 
In contrast, \tool uses multi-head attention and traffic law weighting to create diverse, high-risk violation scenarios, improving test coverage.

\section{Conclusion}\label{sec:Conclusion}
We propose \tool, a novel approach for testing ADS that combines a traffic law-weighted mechanism and a multi-head attention network to generate high-risk, critical scenarios. 
By focusing on high-risk violations and modeling complex traffic interactions, \tool outperformed the state-of-the-art approaches in scenario coverage and scenario diversity.


\bibliographystyle{IEEEtran}
\bibliography{references}

%
%
%
%
%
%
%
%
%

\end{document}